\documentclass[11pt,twoside,onecolumn]{article}
\usepackage[]{latexsym}
\usepackage{epsfig}
\newcommand{\be}{\begin{eqnarray}}
\newcommand{\ee}{\end{eqnarray}}

\title{\bf Probing quantum gravity effects in black holes at LHC}
\author{
G.L.~Alberghi$^{a,b,c}$\thanks{alberghi@bo.infn.it},$\ $
R.~Casadio$^{a,b}$\thanks{casadio@bo.infn.it},$\ $
D.~Galli$^{a,b}$\thanks{galli@bo.infn.it},$\ $
D.~Gregori$^{a,b}$\thanks{gregori@bo.infn.it},$\ $
A.~Tronconi$^{a,b}$\thanks{tronconi@bo.infn.it}$\ $ and 
V.~Vagnoni$^{b}$\thanks{vagnoni@bo.infn.it}
\\
$^a${\em Dipartimento di Fisica, Universit\`a di Bologna}
\\
{\em via Irnerio~46, 40126 Bologna Italy}
\\
$^b${\em Istituto Nazionale di Fisica Nucleare, Sezione di Bologna}
\\
{\em via Irnerio~46, 40126 Bologna, Italy}\\
$^c${\em Dipartimento di Astronomia, Universit\`a di Bologna}
\\
{\em via Ranzani~1, 40126 Bologna Italy}}
\begin{document}
\maketitle
\begin{abstract}
We study modifications of the Hawking emission in the evaporation of
miniature black~holes possibly produced in accelerators when their mass
approaches the fundamental scale of gravity, set to $1\,$TeV according
to some extra dimension models.
Back-reaction and quantum gravity corrections are modelled
by employing modified relations between the black hole mass
and temperature. We release the assumption that black holes explode
at $1\,$TeV or leave a remnant, and let them evaporate to much
smaller masses.
We have implemented such modified decay processes into an existing 
micro-black hole event generator, performing a study of the decay
products in order to search for phenomenological evidence of quantum
gravity effects.
\end{abstract}
\setcounter{page}{1}
\section{Introduction}
\setcounter{equation}{0}
One of the most exciting features of models with large extra
dimensions~\cite{arkani,RS} is that the fundamental scale of gravity
$M_{\rm G}$ could be as low as the electroweak scale
($M_{\rm G}\simeq 1$~TeV) and black holes may therefore
be produced in our accelerators~\cite{banks,Giddings3}.
Black holes have been studied in both compact~\cite{argyres} and
infinitely extended~\cite{chamblin,ehm} extra~dimensions (see
also~\cite{cavaglia} for recent reviews).
The basic feature of black hole production is that its cross section
should essentially be given by the horizon area of the forming black
hole and grows with the centre of mass energy of the colliding particles
as a power which depends on the number of
extra-dimensions~\cite{banks,Giddings3}.
\par
Once the black hole has formed (and after possible transients),
Hawking radiation~\cite{hawking} is expected to set off.
The standard description of this effect leads to the (canonical)
Planckian distribution for emitted particles and black hole life-times
so short that the decay of micro black holes can be viewed
as sudden~\cite{dimopoulos}.
This picture has been implemented in numerical
codes~\cite{charybdis,trunoir} which normally let the black hole decay
down to a mass of order $M_{\rm G}$ via the Hawking law and then
explode into a (selectable) number of decay products.
However, energy conservation is not guaranteed {\em a priori\/}
in the canonical ensemble and the black hole temperature in fact
diverges for vanishing black hole mass.
Energy conservation is therefore enforced by means of kinematical
contraints.
From a theoretical point of view, such issues can be taken care of
by employing the more consistent microcanonical description of black
hole evaporation~\cite{mfd} which has been first applied in the context
of large extra dimensions in Refs.~\cite{bc,hossenfelder1,rizzo}.
It was then shown that actual life-times can vary greatly depending
on the model details~\cite{bcLHC,rizzo}.
Further, the possibility of ending the evaporation by leaving a stable
remnant has also been recently considered~\cite{hossenfelder2}.
\par
The issue of the end of the evaporation is really an open question
theoretically, because we do not have a reliable theory of quantum
gravity.
One could in fact go as far as to say that theoretical physicists
are looking forward to detecting black holes at the Large Hadron
Collider (LHC) in order to finally have experimental data about
quantum gravity.
The very detection of these objects would already be evidence that
$M_{\rm G}$ is not the (naive) Planck energy (about $10^{19}\,$GeV)
and extra dimensions do exist.
Furthermore, the late stage of black hole evaporation (when its mass
$M\sim M_{\rm G}$) could tell us more about the details of
quantum gravity.
The question then naturally arises as whether deviations from the
Hawking law induced by the underlying theory of quantum gravity
can actually be detected.
The purpose of this report is precisely to study if a modified
evaporation law at small black hole mass ($M\sim M_{\rm G}$)
can produce detectable differences provided energy conservation
is properly enforced.
The latter requirement in fact changes the actual distribution
of emitted particles even for the case of the standard Hawking law,
as we shall also show.
For this purpose, we have developed modifications to the CHARYBDIS Monte Carlo
code \cite{charybdis}, which we shall describe in some details in Section~\ref{TTGM}.
Preliminary results from a few test runs are then analyzed in
Section~\ref{sim}.
\par
We use units with $c=\hbar=k_{\rm B}=1$ and denote the fundamental
gravity length in $D$ space-time dimensions as $\ell_{D}$.
\section{Black hole evaporation}
\setcounter{equation}{0}
The no-hair theorems guarantee that a black hole is characterized by
its mass, charges and angular momentum only.
The only parameter characterizing an uncharged non-rotating black hole is
thus its mass $M$.
On considering solutions to the Einstein equations 
(or applying Gauss' theorem) in $4+d$ dimensions, one can derive
the following relation between the mass and the horizon radius of such
a black hole,
\be
R_{\rm H}=\frac{1}{\sqrt{\pi}\,M_{\rm G}}\,
\left(\frac{M}{M_{\rm G}}\right)^{\frac{1}{d+1}}
\left(\frac{8\,\Gamma\left(\frac{d+3}{2}\right)}{d+2}
\right)^{\frac{1}{d+1}}
\ ,
\ee
and its temperature is given by
\be
T_{\rm H} =\frac{d+1}{4\,\pi\,R_{\rm H}}
\ .
\label{TH}
\ee
\par
Once formed, the black hole begins to evolve.
In the standard picture the evaporation process can be 
divided into three characteristic stages~\cite{Giddings3}:
\begin{enumerate}
\item {\sc Balding phase:}
the black hole radiates away the multipole moments
it has inherited from the initial configuration,
and settles down in a hairless state.
A certain fraction of the initial mass will also be
lost in gravitational radiation.
\item {\sc Evaporation phase:}
it starts with a spin down phase in which the Hawking
radiation~\cite{hawking} carries away the angular momentum, after
which it proceeds with the emission of thermally distributed
quanta until the black hole reaches the Planck mass (replaced by
the fundamental scale $M_{\rm G}$ in the models we are considering
here).
The radiation spectrum contains all the Standard Model particles,
which are emitted on our brane, as well as gravitons, which are
also emitted into the extra~dimensions.
It is in fact expected that most of the initial energy is emitted
during this phase into Standard Model particles~\cite{ehm}
although this conclusion is still being debated (see,
e.g.~Ref.~\cite{cavaglia2}).
\item {\sc Planck phase:}
once the black hole has reached a mass close to the effective Planck
scale $M_{\rm G}$, it falls into the regime of quantum gravity and
predictions become increasingly difficult.
It is generally assumed that the black hole will either completely
decay in some last few Standard Model particles or a stable remnant
be left which carries away the remaining energy~\cite{hossenfelder2}.
\end{enumerate}
\par
In our approach we will consider possible modifications to the
second and third phases.
On one hand, we will modify the Hawking phase by employing a
different relation between the horizon radius and the temperature.
On the other hand, we will look at the possibility that the evaporation
may not end at the fundamental scale $M_{\rm G}$ ($\sim 1\,$TeV)
but proceeds further until a lower energy has been reached.
\subsection{Quantum gravity and Monte Carlo generator}
\label{TTGM}
Several Monte Carlo codes which
simulate the production and decay of micro-black holes
(see, e.g.~Refs.~\cite{charybdis,trunoir,ahn}) are available nowadays.
Instead of developing from scratch a new event generator, we have
found convenient to adopt the CHARYBDIS code, implementing relevant
modifications therein. In the following we are going to describe
them in some detail.
\par
In order to let the black hole evaporate below
$M_{\rm G}$, one needs to treat the mass of the emitted particles
properly.
That includes having the correct phase space measure in the
Planckian number density for the emitted particles,
\be
N_m(\vec k)=\frac{d^3 k}{e^{\omega/T_{\rm H}}\pm 1}
\ ,
\label{N_k}
\ee
where $m$ is the particle mass, $\vec k$ the particle 3-momentum
and $\omega=\sqrt{k^2+m^2}$.
Moreover, since $N_m$ depends on $m$ and not just on the statistics,
one cannot assume a fixed ratio of production for fermions versus
bosons but particle multiplicity should be used when generating
particle types randomly.
We therefore use the multiplicities predicted by the Standard
Model~\cite{pdg}.
\par
\begin{figure}[t]
\centerline{
\raisebox{4cm}{$T$}
\epsfxsize=200pt\epsfbox{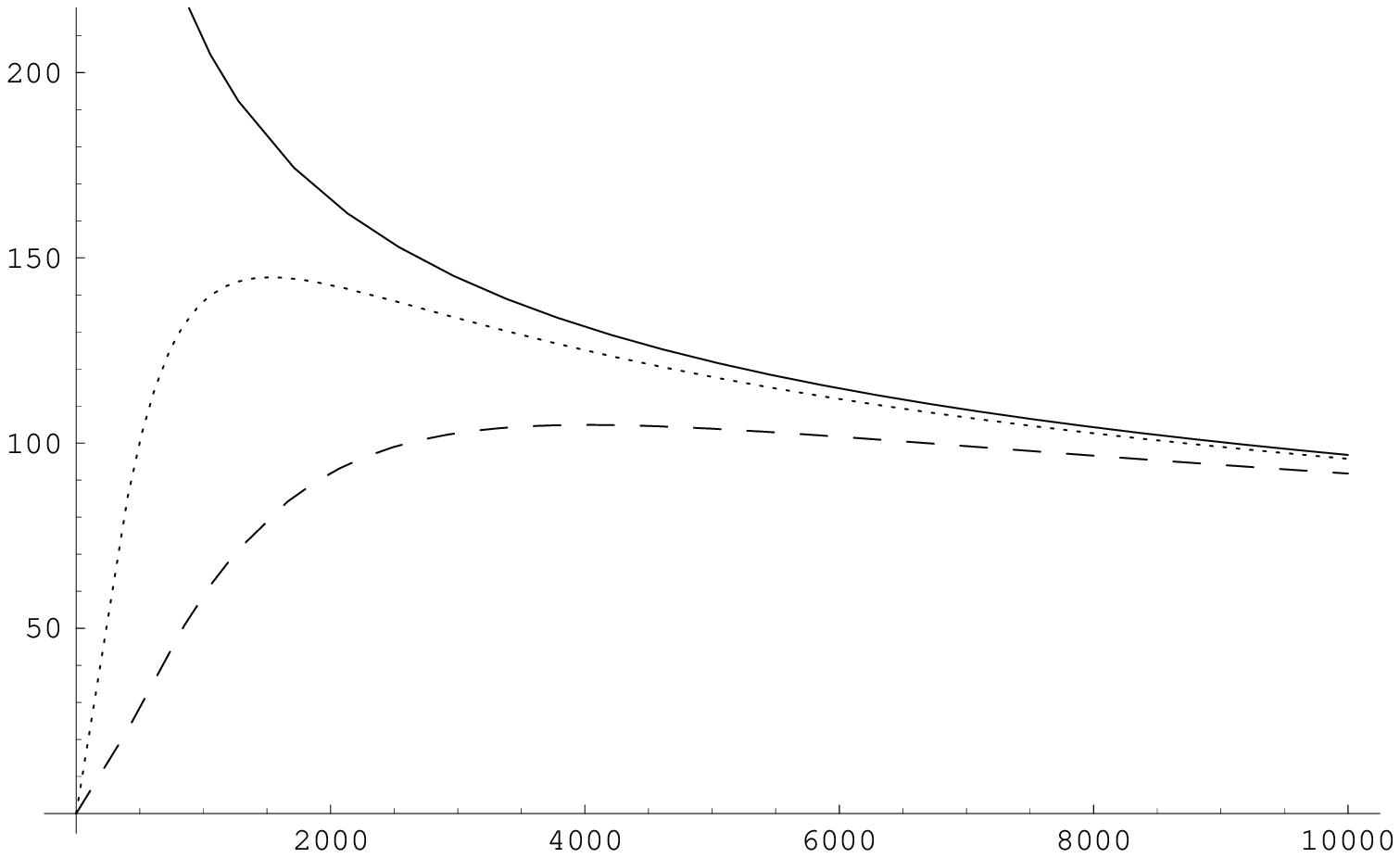}
\hspace{0.5cm}
\raisebox{4cm}{$T$}
\epsfxsize=200pt\epsfbox{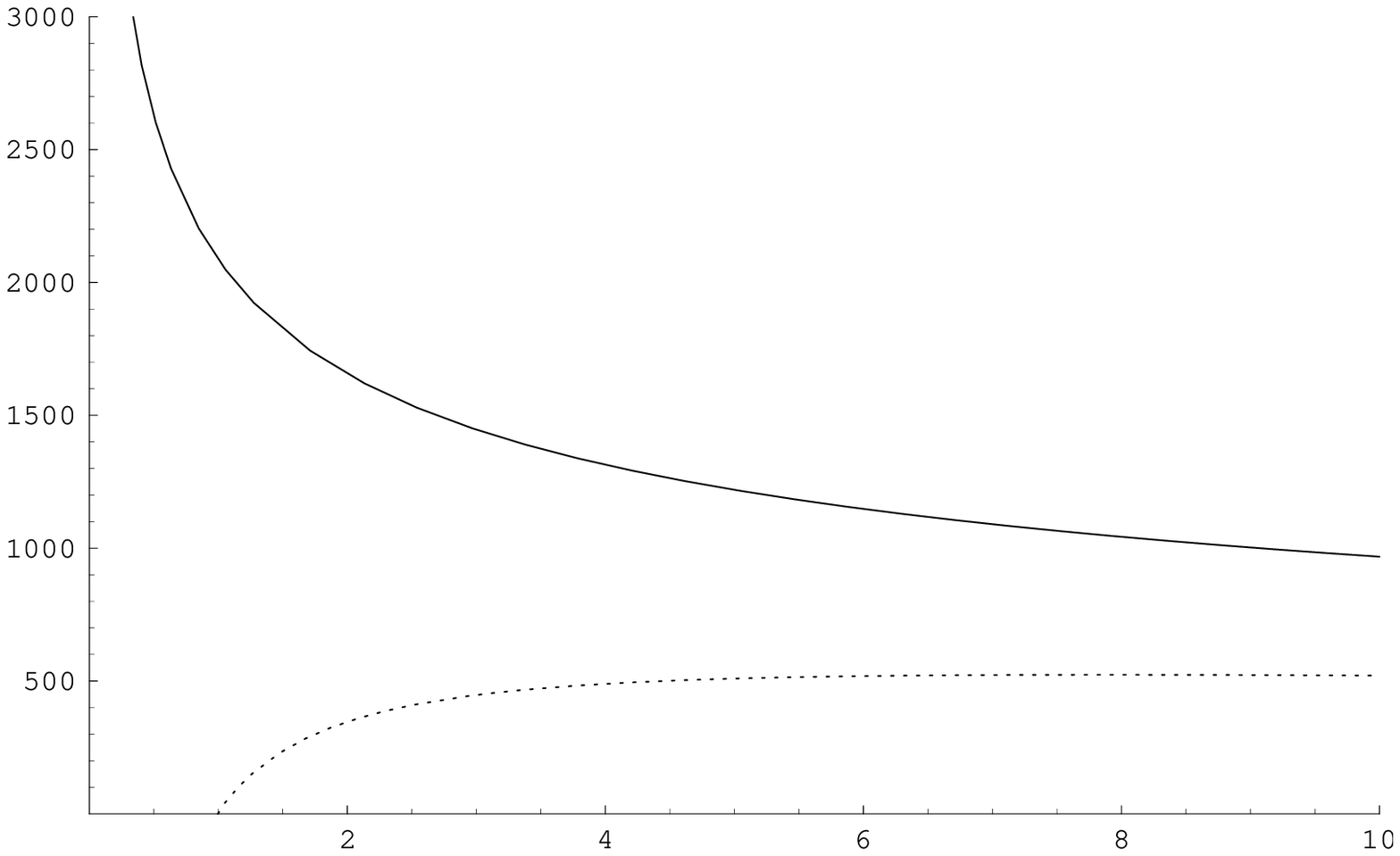}
}
\hspace{7cm}
$M$
\hspace{7cm}
$M$ 
\caption{Black hole temperature (in GeV) as a function of the black hole
mass (in GeV).
In the left panel, we compare the Hawking law (solid line) to
the case in Eq.~(\ref{Tmod1}) for $\ell=10^{-3}\,$GeV$^{-1}$, $\alpha=5$, $n=5$
(dashed line) and $\ell=10^{-3}\,$GeV$^{-1}$, $\alpha=5$, $n=1$ (dotted line).
In the right panel, the Hawking law (solid line) is compared
to the modified case~(\ref{Tmod2}) for $\ell=1.14\cdot 10^{-4}\,$GeV$^{-1}$,
$n_1=1$, $n_2=1$ (dotted line).
\label{T}
}
\end{figure}

In order to include possible quantum gravity effects, we then
employ modified expressions for the temperature of the form
\be
T=F_i\,T_{\rm H}
\ ,
\ee
in which, to cover results in the existing literature (for a partial
list of approaches to the problem, see
Refs.~\cite{mfd,hossenfelder1,rizzo,bcLHC,nicolini}), we shall
consider two possible modifying factors,
\be
F_{1}(\ell,\alpha,n)=\frac{R_{\rm H}^n}{R_{\rm H}^n+\alpha\,\ell^n}
\label{Tmod1}
\ee
and
\be
F_{2}(\ell,n_1,n_2)=\left[1-\left(\frac{\ell}{R_H}\right)^{n_1}\right]^{n_2}
\label{Tmod2}
\ ,
\ee
where $\ell$, $\alpha$, $n$, $n_1$, and $n_2$ are parameters that can be
adjusted (see below).
Note that $F_1$ leads to a vanishing temperature for vanishing black hole
mass (horizon radius), whereas with $F_2$ the temperature vanishes at
finite $M$ (see~\cite{nicolini} and References therein).
Specific examples are shown in Fig.~\ref{T} together with the
standard Hawking law~(\ref{TH}).
\par
A list of some adjustable parameters is given in Table~\ref{FreePar} and
we just wish to make a few remarks.
The initial black hole mass $M_0$ can be either fixed (to within the maximum
centre mass energy expected at the LHC) or generated randomly.
During the decay, when $M<M_{\rm f}$ the black hole explodes in a selectable
number $N_{\rm f}$ of fragments.
An important aspect of our analysis is that we set the grey-body factors
to 1 for all kinds of particles.
It is to be expected that this approximation becomes too restrictive,
particularly if the final black hole mass is of the order of (a few) GeV.
However, we are not currently aware of any computations of grey-body factors
which include massive particles and have left this issue for future developments.
\begin{table}[h]
\centerline{
\begin{tabular}{|c|c|c|}
\hline
$M_0$
&
Initial black hole mass
&
$1-14$\,TeV; random
\\
\hline
$M_{\rm f}$
&
Minimum black hole mass
&
$1-1000$\,GeV
\\
\hline
$N_{\rm f}$
&
number of final fragments
&
$2-6$
\\
\hline
$d$
&
number of extra dimensions
&
$2-6$
\\
\hline
$F_i$
&
modified temperature
&
$1$ (no modification), $F_1$, $F_2$
\\
\hline
\end{tabular}
}
\caption{Parameters that can be adjusted in the Monte Carlo generator
and their ranges.
For each modified temperature, one can also set the corresponding
parameters as described in the text.}
\label{FreePar}
\end{table}
\section{Simulation results}
\label{sim}
\begin{figure}[ht]
\centerline{
\epsfxsize=220pt\epsfbox{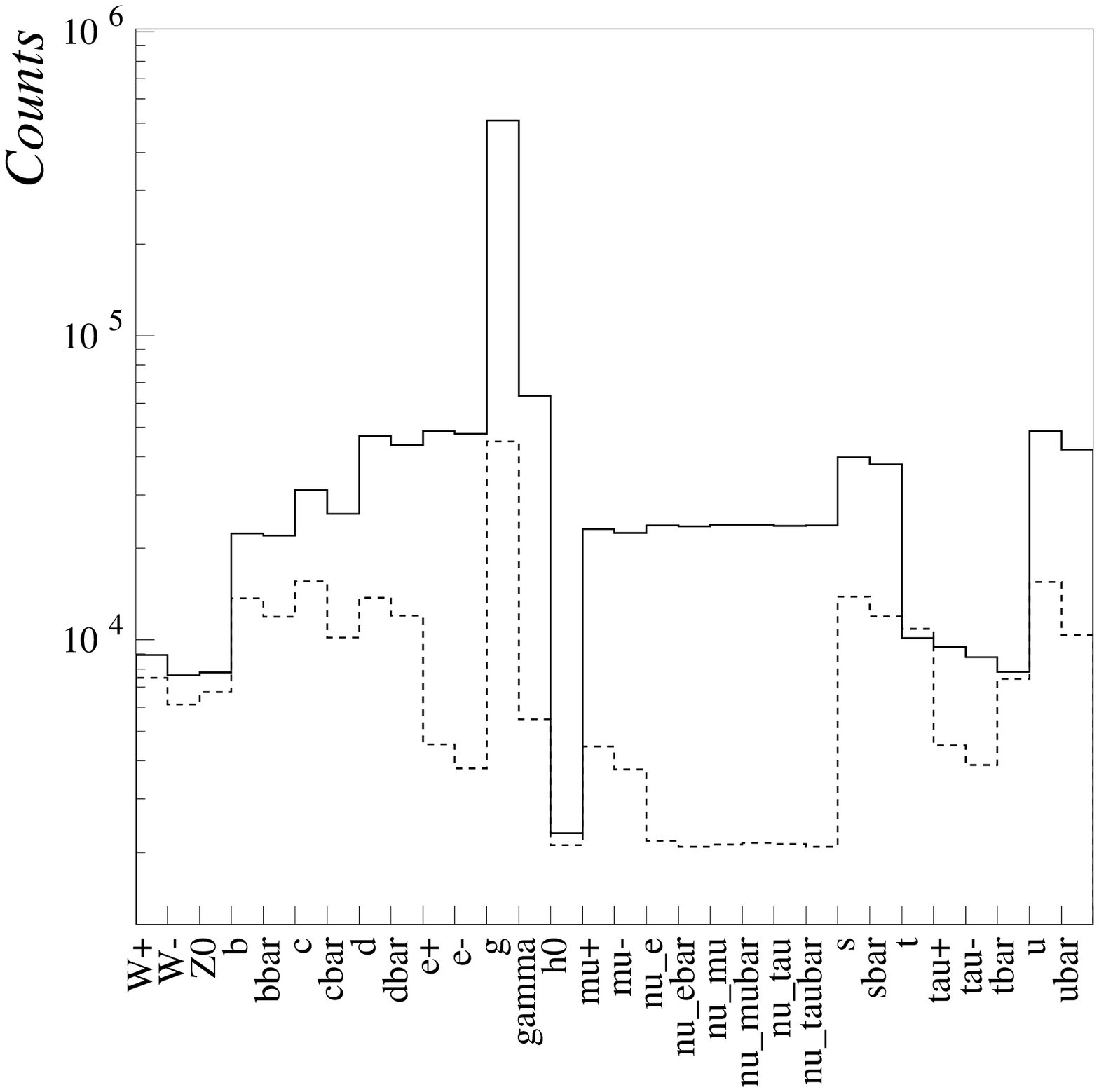}
\hspace{0.5cm}
\epsfxsize=220pt\epsfbox{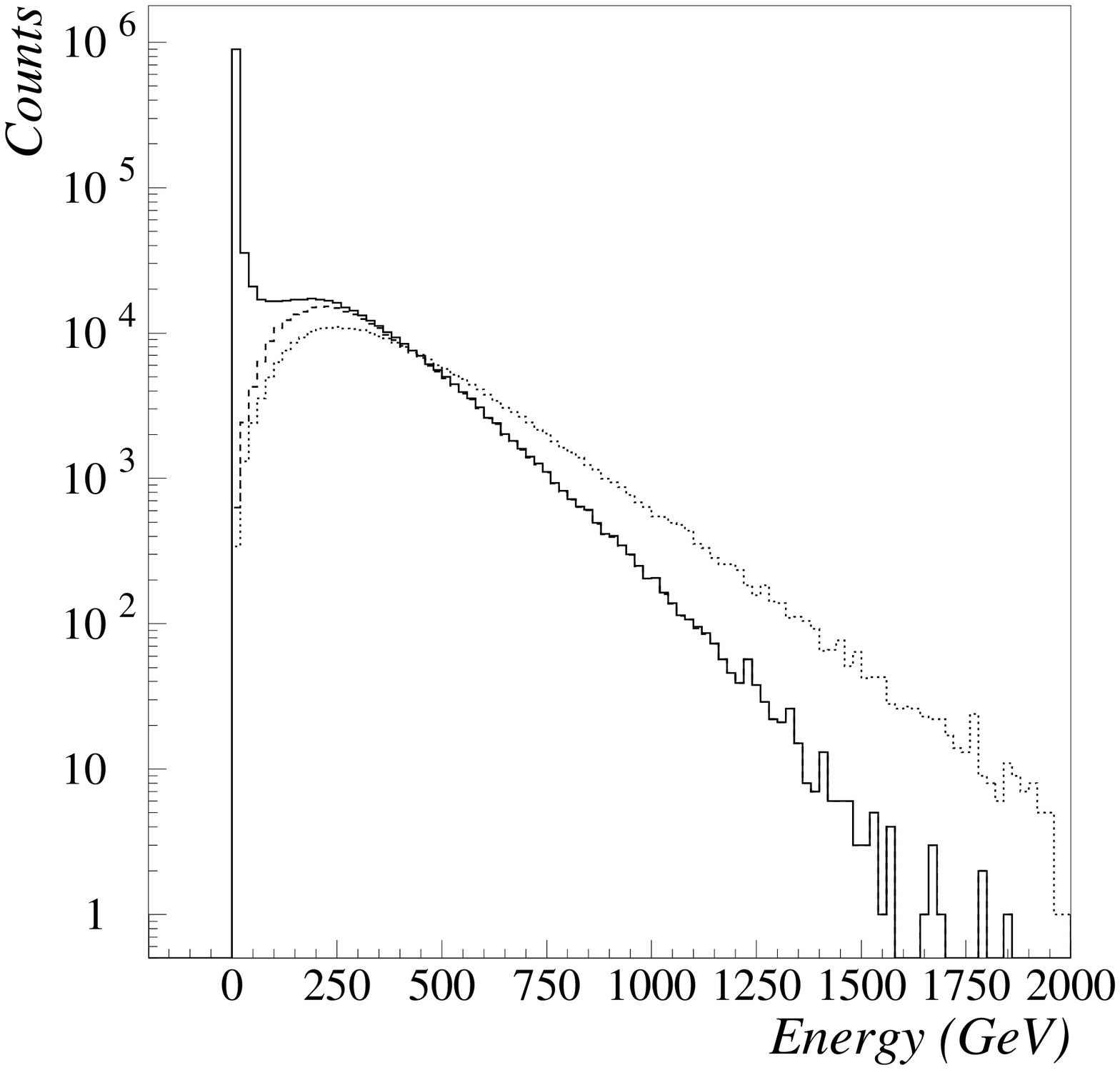}
}
\caption{Left panel: abundance of Standard Model particles produced during
the evaporation of 10~TeV black holes, for a run of 10000 events with
the modified law~(\ref{Tmod1}) (solid histogram) and with the standard
CHARYBDIS generator (dashed histogram).
The fundamental scale of gravity is set to 1~TeV, temperature parameters
are the same as in left panel of Fig.~\ref{T} and number of extra dimensions
$d=2$.
Right panel: energy distribution of the emitted particles obtained
with the modified law~(\ref{Tmod1}) (solid histogram), including only
particles emitted when the black hole has a mass exceeding 1~TeV
(dashed histogram), and with the standard CHARYBDIS generator
(dotted histogram).
\label{panam}
}
\end{figure}
The standard CHARYBDIS generator simulates the evaporation of a micro-black
hole according to the Hawking's law until the black hole mass reaches the
fundamental scale of gravity $M_{\rm G}$ in a given model of extra dimensions
(e.g.,~at 1~TeV).
Once the scale $M_{\rm G}$ is reached, the black hole remnant is decayed into
a few bodies simply according to phase space.
As described in the previous section, in our modified version we implemented
ways to cope with the evaporation well below the fundamental scale, down to
negligible black hole masses.
We shall thus compare the outcomes of the generator for the standard and
modified versions from a phenomenological point of view. 
\par
In the simulations run this far, we noted that modified
temperatures~(\ref{Tmod1}) and~(\ref{Tmod2}) produced qualitatively similar
results when the black hole mass such that $R_{\rm H}>\ell$.
Further, the case of Eq.~(\ref{Tmod2}) corresponds to a remnant of radius
$R_{\rm H}=\ell$ which has been considered elsewhere~\cite{hossenfelder2}.
We shall therefore refer only to the results obtained for the temperature modified
by the factor $F_1$ with the parameters given in the caption of Fig.~\ref{T}.
Since the temperature behaviour at low mass values is no longer divergent
and tends to zero for $M=0$ in this case, we do expect a continuous
emmission of increasingly softer particles until the black hole evaporates
completely.
Once the black hole mass has decreased below the threshold for
producing a given particle, such a particle can no longer be emitted.
Hence one expects that the production of the heavier particles, i.e.~massive gauge
bosons and top quarks, is scarsely affected, while the emission of soft
low mass (or massless) particles is largely enhanced.
\par
A comparison of the relative abundance of the Standard Model particles
produced by the black hole evaporation with and without using the modified
law of the form~(\ref{Tmod1}) -- before parton evolution and hadronization are
performed -- is shown in the left panel of Fig.~\ref{panam}.
In the right panel of Fig.~\ref{panam} we show the energy distribution
of the emitted particles.
As expected, the modified law dramatically changes the spectrum
at low energy, leaving the spectrum at large energy moderately affected.
This translates into a much larger multiplicity of isotropically emitted
soft particles which might be a possible phenomenological signature of
quantum gravity effects.
\section{Conclusions}
\setcounter{equation}{0}
We have considered possible modifications for the decay of
micro-black holes around the fundamental scale of gravity
$M_{\rm G}$ (of order $1$~TeV).
Inspired by the microcanonical description of the Hawking
evaporation and other treatments in the literature, we have
studied modified statistical laws for the temperature 
together with the requirement of energy conservation enforced
up to the end of the evaporation at a scale
$M_{\rm f}\ll M_{\rm G}$.
\par
The numerical simulations we have run so far suggest that
such modification mainly affect the total number of light and massless
particles emitted at very small energy.
On lowering the minimum black hole mass $M_{\rm f}$ increases
this number. The detection of the isotropic emission of a large number of
soft isolated particles (i.e. not associated to jet-like topologies)
could be an evidence in favour of a specific decay law or of a
specific theory of quantum gravity.
\par
In our analysis we have not explicitly considered the case in which
black holes leave stable remnants since it has already been
reported in Ref.~\cite{hossenfelder2}.
Let us just mention that very large values of $\alpha$ and/or $\ell$
in the modified temperature~(\ref{Tmod1}) (together with energy conservation)
would make it very hard for the black holes to continue the emission for
$M\ll M_{\rm G}$ and stable remnants would be effectively produced. 
\end{document}